\documentclass[aps,reprint,twocolumn,superscriptaddress]{revtex4-1}
\usepackage{subfigure}
\usepackage{amsmath, amssymb,graphicx,color,bm,epstopdf}
\usepackage[dvipsnames]{xcolor}
\usepackage{bbold}
\usepackage{braket}
\usepackage{comment}
\usepackage{natbib}
\usepackage{mathtools}

\newcommand{\be}{\begin{equation}}
\newcommand{\e}{\end{equation}}
\newcommand{\beml}{\begin{subequations}}
\newcommand{\eml}{\end{subequations}}
\newcommand{\beq}{\begin{eqnarray}}
\newcommand{\eq}{\end{eqnarray}}
\newcommand{\ba}{\begin{array}}
\newcommand{\ea}{\end{array}}
\newcommand{\bpm}{\begin{pmatrix}}
\newcommand{\epm}{\end{pmatrix}}
\newcommand{\bc}{\begin{cases}}
\newcommand{\ec}{\end{cases}}

\usepackage{placeins}
\usepackage{float}

\definecolor{amendments}{rgb}{0.0, 0.0, 0.7}

\usepackage[utf8]{inputenc}
\begin{document}
\title{Pair production seeded by electrons in noble gases\\ as a method for the laser intensity diagnostics}
\author{I.~A. Aleksandrov}
\email{i.aleksandrov@spbu.ru}
\affiliation{Department of Physics, St. Petersburg State University, Universitetskaya Naberezhnaya 7/9, Saint Petersburg 199034, Russia}
\affiliation{Ioffe Institute, Politekhnicheskaya str. 26, Saint Petersburg 194021, Russia}
\author{A.~A. Andreev}
\affiliation{Department of Physics, St. Petersburg State University, Universitetskaya Naberezhnaya 7/9, Saint Petersburg 199034, Russia}
\affiliation{ELI-ALPS, ELI-HU NKft. Dugonics t\'{e}r 13, Szeged 6720, Hungary}
\affiliation{Max-Born Institut, Max-Born Str. 2A, Berlin 12489, Germany}

\begin{abstract}
In this study we explore the possibility of using the process of electron-positron pair creation in strong laser fields as a tool for measuring the intensity of the corresponding laser radiation. In the initial state we consider either free electron gas or gas of neutral xenon, the electrons of which get ionized. Once these seed electrons gain sufficient energy in the external laser field, they can emit high-energy photons which subsequently decay producing electron-positron pairs via the Breit-Wheeler mechanism. By detecting the resulting positrons, one can recover the value of the laser intensity by means of the one-to-one correspondences deduced in the present investigation. We analyze two different configurations of the external field: the setup involving an individual focused laser pulse and the combination of two counterpropagating laser pulses. Performing numerical calculations and analyzing their accuracy, we demonstrate that based on our estimates, the laser intensity can be determined within the range $10^{23}$--$10^{26}~\text{W}/\text{cm}^2$ with a relative uncertainty of $10$--$50 \%$.
\end{abstract}

\maketitle

\section{Introduction}

The current and forthcoming laser facilities should allow one to achieve a laser intensity of $10^{22}~\text{W}/\text{cm}^2$ or even higher paving the way for probing various strong-field QED phenomena (see, e.g., Refs.~\cite{vulkan_2010, bashinov_epj_2014, zou_2015, kawanaka_2016, papadopoulos_2016, weber_mre_2017, gales_2018, guo_2018, danson_2019, yoon_2019}). Nevertheless, the determination of the peak intensity of such strong laser pulses represents a formidable task. Recently, a number of various techniques have been extensively discussed in the literature: measuring yields of highly charged ions due to atomic ionization~\cite{ciappina_pra_2019, ciappina_lpl_2020, ciappina_mra_2020}, detecting the light scattering or additional radiation due to the interaction between electrons and the laser field~\cite{har-shemesh_2012, Vais_2016, yan_2017, harvey_2018, he_2019, blackburn_2020}, or the analysis of photoionization or direct acceleration of charged particles~\cite{kalashnikov_2015, vais_2017, vais_2018, ivanov_2018, krajewska_2019, mackenroth_2019, vais_2020, vais_2021} (see also references therein). In the present paper we discuss how the laser intensity diagnostics can be carried out using the strong-field QED mechanism of pair production due to the interaction of free electrons or free xenon atoms with an intense laser field. As this scenario reflects a threshold dependence on the field amplitude, it can allow one to accurately resolve the laser intensity by measuring the positron yield.

We examine the following setup. A high-intensity linearly polarized laser pulse (or two counterpropagating pulses) travels through a gas chamber with neutral xenon whose electrons after ionization serve as seed particles for the subsequent process of nonlinear Compton scattering providing high-energy photons. These photons can then decay via the Breit-Wheeler mechanism yielding electron-positron pairs which can cascade further. As this process of positron production possesses a quite sharp threshold, measuring the positrons produced provides an accurate method for the determination of the laser intensity. We will show that the intensity can be extracted within the domain $10^{23}~\text{W}/\text{cm}^2 \lesssim I \lesssim 10^{26}~\text{W}/\text{cm}^2$ depending on the laser field configuration. Besides, we also consider free electrons instead of xenon atoms in order to demonstrate the advantage of the latter scenario due to more efficient acceleration of the inner-shell electrons~\cite{maltsev_prl_2003, hu_pre_2006, karmakar_2007, artemenko_pra_2017}. We also note that positron production via the Breit-Wheeler process was considered as a method for the intensity diagnostics in Refs.~\cite{lecz_andreev_2019, lecz_andreev_2020}, where the laser field was focused on the surface of a thin foil instead of traveling through a gas chamber.

In this study we use a relatively simple method for estimating the total number of positrons created without performing Monte-Carlo simulations, which have been used in numerous studies (see, e.g., Refs.~\cite{artemenko_pra_2017, lecz_andreev_2019, lecz_andreev_2020, nerush_prl_2011, elkina_2011, king_pra_2013, tamburini_srep_2017, vranic_sci_rep_2018, baumann_sci_rep_2019, blackburn_njp_2019, gu_2019, yakimenko_2019}). In what follows, we will successively take into account ionization, acceleration in the laser field, photon emission, and Breit-Wheeler pair production neglecting further stages of the cascade where the particles created emit new photons. To make sure that we receive significantly accurate predictions for the total number of positrons, we will not consider too intense laser pulses which initiate an avalanche-like reaction~\cite{fedotov_prl_2010, 2015_ufn_narozh}. It turns out that even if one disregards the further cascading, our scheme can be utilized in a quite large intensity interval.

The paper has the following structure. In Sec.~\ref{sec:ionization} we discuss how the ionization process is described within our numerical simulations. In Sec.~\ref{sec:dynamics} we briefly present the external field configuration and the main points of how the electron dynamics is computed. In Sec.~\ref{sec:pairs} we describe how the QED mechanisms of photon emission and Breit-Wheeler pair production are incorporated in our calculations. In Sec.~\ref{sec:effects} we discuss and estimate other effects which are not taken into account in our numerical procedures. In Sec.~\ref{sec:results} we present and discuss the results obtained. Finally, we conclude in Sec.~\ref{sec:discussion}.

We will use atomic units: Planck constant $\hbar=1$, electron mass $m=1$, electron charge $e=-1$. In these units the speed of light in vacuum is $1/\alpha\approx 137.036$, where $\alpha$ is the fine-structure constant.

\section{Ionization model}\label{sec:ionization}

Atoms interacting with a strong laser background get ionized providing seed electrons for subsequent emission of high-energy photons which afterwards yield positrons via the Breit-Wheeler mechanism. Here we discuss how one can describe the ionization stage. Consider an atom at a given position $\boldsymbol{r}_0 = (x_0, y_0, z_0)$. Although it interacts with the laser pulse, we assume it to be motionless since in what follows we will sum the results over $\boldsymbol{r}_0$ within the interaction region, where the atoms are distributed randomly. For this reason we also assume that after ionization the free electrons appear at rest at the same point $\boldsymbol{r}_0$. First, we evolve in time the ionization probabilities $P_j$ for each electron level $j$ (in Xe there are 54 of them) via
\begin{equation}
dP_j(t) = [1 - P_j(t)]W_j(t)dt,\quad j = 1, 2, ..., N_\text{lev},
\label{eq:P_ioniz}
\end{equation}
where $W_j(t)$ is the ionization rate depending on $j$, the effective charge $Z_j (t)$, and the electric field strength $E(t) \equiv E(\boldsymbol{r}_0, t)$. At the time instant $t = t_\text{in}$, when the atom starts to interact with the laser field, we assume that $Z_j = j$ and at each time step, we recalculate these charges according to
\begin{equation}
Z_j(t) = j + \sum_{k=j+1}^{N_\text{lev}} P_k(t),
\label{eq:Z_recalc}
\end{equation}
i.e., we take into account that the screening of the nuclear charge reduces as the electrons get ionized. The main problem is to accurately evaluate the ionization rate $W(t)$. Hereinafter we omit $j$ considering a given energy level. Let $I_p$ and $l$ denote the corresponding ionization potential and orbital quantum number, respectively (they are taken from the NIST database~\cite{nist_ie}). We also introduce the following notations:
\begin{eqnarray}
\kappa &=& \sqrt{2I_p},\label{eq:not_kappa}\\
\nu &=& Z/\kappa, \label{eq:not_nu}\\
\mathcal{F}(t) &=& |E(t)|/\kappa^3.
\label{eq:not_F}
\end{eqnarray}
Since the external laser field is extremely strong, the corresponding Keldysh parameter $\gamma_\text{K} = \sqrt{2I_p}\omega/|E|$ is very small, $\gamma_\text{K} \lesssim 10^{-3}$. It means that one can employ the closed-form expressions derived for tunnel ionization.

The simplest model is the so-called Ammosov-Delone-Krainov (ADK) model~\cite{ammosov_1986}:
\begin{equation}
W_\text{ADK} (t)  = \frac{|E(t)|}{8\pi Z} \, \sqrt{\frac{3\mathcal{F}(t)}{\pi}} \bigg ( \frac{4 \mathrm{e}}{\nu \mathcal{F}(t)} \bigg )^{2\nu} \mathrm{exp} \bigg (-\frac{2}{3\mathcal{F}(t)} \bigg ).
\label{eq:ADK}
\end{equation}
The WKB approach used in the derivation of Eq.~\eqref{eq:ADK} is not justified if the ionization potential is sufficiently small, which leads to the so-called barrier-suppression (BS) effects extensively discussed in the literature (see, e.g., Refs.~\cite{ciappina_lasphyslett_2020, artemenko_pra_2017, tong_jpb_2005, krainov_josab_1997, zhang_pra_2014}). The characteristic BS field reads~\cite{tong_jpb_2005}
\begin{equation}
\mathcal{F}_\text{BS} = \frac{1}{16 \nu}.
\label{eq:F_BS}
\end{equation}
To take into account the BS effects, one can use an empirical formula proposed in Ref.~\cite{tong_jpb_2005} multiplying Eq.~\eqref{eq:ADK} by the factor $\mathrm{exp} [-2 \overline{\alpha} \nu^2 \mathcal{F}(t)]$ for $\mathcal{F}(t) \gtrsim \mathcal{F}_\text{BS}$. (for xenon $\overline{\alpha} = -9.0$). However, this approximate correction yields relatively accurate results only when $\mathcal{F} \lesssim 2 \mathcal{F}_\text{BS}$~\cite{zhang_pra_2014}. In Ref.~\cite{zhang_pra_2014} the authors proposed a more sophisticated correction which is valid up to $\mathcal{F} \approx 4.5 \mathcal{F}_\text{BS}$. Nevertheless, in our computations it is still a very limited domain. To go beyond this region, i.e., consider $\mathcal{F} \gg \mathcal{F}_\text{BS}$, we will employ the approach presented in Ref.~\cite{artemenko_pra_2017}. Using the ADK model, we compute the ionization rates according to
\begin{equation}
W (t) = \begin{dcases}
W_\text{ADK} (t) & \text{if}~\mathcal{F} < \mathcal{F}_\text{BS},\\
W_\text{ADK} \, \mathrm{exp} [-2 \overline{\alpha} \nu^2 \mathcal{F}(t)] (t) & \text{if}~\mathcal{F}_\text{BS} \leq \mathcal{F}(t) \leq \mathcal{F}_\text{L},\\
W_\text{L} (t) & \text{if}~\mathcal{F} > \mathcal{F}_\text{L},
\end{dcases}
\label{eq:ioniz_prob_full}
\end{equation}
where $W_\text{L} (t) = 2I_p \mathcal{F}(t)$ and $\mathcal{F}_\text{L}$ is chosen, so that the function~\eqref{eq:ioniz_prob_full} is continuous.

Having evaluated the functions $P_j (t)$, we then calculate the total number of electrons ionized $N^{(\text{el})} (t) = \sum_j P_j (t)$ and split the $t$ axis into small intervals $[t_i, t_i + \Delta t]$. Each ``portion of electrons'' $\Delta N^{(\text{el})}_i = N^{(\text{el})} (t_i + \Delta t) - N^{(\text{el})} (t_i)$ propagates then according to the relativistic equations of motion. In other words, these classical portions (``macroparticles'') are treated within the particle-in-cell (PIC) approach~\cite{hockney_1988,birdsall_1991}. They give rise to photon emission with the corresponding weights $\Delta N^{(\text{el})}_i$. In the case of free electrons, we deal with only one electron portion of unit weight, so the ionization stage is omitted.

Finally, we note that we also employed the Perelomov-Popov-Terent'ev (PPT) model~\cite{ppt_1966, ppt_1967} instead of the ADK model and obtained similar results. The corresponding slight discrepancies were used in estimating the final uncertainties presented in Sec.~\ref{sec:results}.

\section{Electron dynamics and the laser field configuration}\label{sec:dynamics}

Assuming the electron to be a classical relativistic particle traveling in arbitrary electric and magnetic fields $\boldsymbol{E} = \boldsymbol{E} (t, \boldsymbol{r})$ and $\boldsymbol{H} = \boldsymbol{H} (t, \boldsymbol{r})$, one has to solve the following system of equations:
\begin{eqnarray}
\frac{d\boldsymbol{p}}{dt} &=& -\boldsymbol{E} - \frac{1}{c} \, \boldsymbol{v} \times \boldsymbol{H} + \boldsymbol{F}_\text{rec},\label{eq:eom}\\
\frac{d\boldsymbol{r}}{dt} &=& \boldsymbol{v}, \label{eq:r-v}\\
\boldsymbol{v} &=& \frac{\boldsymbol{p}}{\sqrt{1+\boldsymbol{p}^2/c^2}}. \label{eq:v-p}
\end{eqnarray}
Here $\boldsymbol{F}_\text{rec}$ is the recoil force which appears due to photon emission and will be specified in the next section. The initial conditions have the form $\boldsymbol{p} (t_i) = 0$, $\boldsymbol{r} (t_i) = \boldsymbol{r}_0$, where $t_i$ is the time instant when the electron portion under consideration gets ionized ($t_i > t_\text{in}$). Once the Lorentz parameter $\gamma (t) = \sqrt{1 + \boldsymbol{p}^2 (t) /c^2}$ becomes large ($\gamma \gg 1$), the electron can emit a high-energy photon via nonlinear Compton scattering.

Let us now present the explicit form of the external background in the case of an individual laser pulse propagating along the $z$ axis. The external field is assumed to be a focused Gaussian beam multiplied by a spatial envelope function within the paraxial approximation (see, e.g., Ref.~\cite{Vais_2016}):
\begin{eqnarray}
H_x (t, \boldsymbol{r}) &=& 0, \label{eq:field_Hx}\\
H_y (t, \boldsymbol{r}) &=& \frac{E_0 \rho_0}{\rho (z)} \, f(\omega t - k_0 z) \notag \\
{} & \times & \mathrm{exp} \Big [ -\frac{x^2 + y^2}{2 \rho_0^2} \Big ] \sin \phi, \label{eq:field_Hy}\\
H_z (t, \boldsymbol{r}) &=& \frac{E_0 \rho_0 y}{k_0 \rho_\text{F} \rho^2 (z)} \, f(\omega t - k_0 z) \notag \\
{} & \times & \mathrm{exp} \Big [ -\frac{x^2 + y^2}{2 \rho_0^2} \Big ] \cos \tilde{\phi}, \label{eq:field_Hz}\\
E_x (t, \boldsymbol{r}) &=& H_y (t, \boldsymbol{r}), \label{eq:field_Ex}\\
E_y (t, \boldsymbol{r}) &=& 0, \label{eq:field_Ey}\\
E_z (t, \boldsymbol{r}) &=& \frac{E_0 \rho_0 x}{k_0 \rho_\text{F} \rho^2 (z)} \, f(\omega t - k_0 z) \notag \\
{} & \times & \mathrm{exp} \Big [ -\frac{x^2 + y^2}{2 \rho_0^2} \Big ] \cos \tilde{\phi}, \label{eq:field_Ez}
\end{eqnarray}
where $k_0 \equiv 2\pi/\lambda$, $\rho (z) = \rho_0 \sqrt{(1 - z/\mathfrak{F})^2 + (z/z_*)^2}$,
\begin{eqnarray}
\phi (t, \boldsymbol{r}) &=& \omega t - k_0 z + \arctan \frac{z\mathfrak{F}}{z_* (\mathfrak{F}-z)} + \pi \theta (z-\mathfrak{F}) \notag \\
{} &-& \frac{(x^2 + y^2) [z + (z_*/\mathfrak{F})^2 (z - \mathfrak{F})]}{zz_* \rho^2 (z)} - \varphi, \label{eq:field_phi}\\
\tilde{\phi} (t, \boldsymbol{r}) &=& \phi (t, \boldsymbol{r}) + \arctan \frac{z\mathfrak{F}}{z_* (\mathfrak{F}-z)} + \pi \theta (z-\mathfrak{F}) \notag \\
{} &-& \arctan \frac{z_*}{\mathfrak{F}}, \label{eq:field_phi_tilde}
\end{eqnarray}
and $f(\xi)$ is a smooth envelope function containing $N_\text{c}$ carrier cycles, $f(\xi) = \sin^2[\xi/(2N_\text{c})]\theta(\pi N_\text{c} - |\xi -\pi N_\text{c}|)$. To specify the external field configuration, one has to define two more parameters besides the field amplitude $E_0$, wavelength $\lambda$, number of the carrier cycles $N_\text{c}$, and the CEP parameter $\varphi$. These can be, for example, focal spot radius $\rho_\text{F}$ and waist position $z_\text{F}$. Using these quantities, one derives the rest parameters:
\begin{eqnarray}
\rho_0 &=& \rho_\text{F} \sqrt{1 + [z_\text{F}/(k_0 \rho^2_\text{F})]^2},\label{eq:rho_0}\\
\mathfrak{F} &=& z_\text{F} \bigg ( 1 + \frac{k_0^2 \rho^4_\text{F}}{z^2_\text{F}} \bigg ), \label{eq:F} \\
z_* &=& k_0 \rho_0^2. \label{eq:zstar}
\end{eqnarray}
In what follows, we assume $\lambda = 1.0~\mu\text{m}$, $\rho_\text{F} = 2.0~\mu\text{m}$, $z_\text{F} = 20.3~\mu\text{m}$, and $\varphi = 0$, which leads to $\mathfrak{F} = 51.4~\mu\text{m}$ and $f_\# \equiv \mathfrak{F}/(2 \rho_0) = 10.0$. The beam divergence $\Delta = \arctan~(k_0 \rho_0)^{-1} \approx 0.06$. The number of the carrier cycles is $N_\text{c} = 10$, so the pulse duration is $\tau = \lambda N_\text{c} / c = 33.3~\text{fs}$.

Besides a single laser pulse, we will also analyze the field configuration consisting of two focused pulses crossing at some angle $\theta$. In principle, our numerical simulations can be carried out for arbitrary values of $\theta$, but we will focus on the extreme case $\theta = \pi$ of two counterpropagating pulses. By means of both the single-pulse case and this setup, one covers the largest domain of the laser intensities that can be measured by detecting the resulting positrons. Using intermediate values of $\theta$ may be more profitable to work within various subintervals as it allows one to tune the field configuration to make the sharp threshold behavior of the pair-production process coincide with the intensity region of interest thus reducing the corresponding uncertainties. We also note that introducing a high-energy electron beam instead of almost restless electron (xenon) gas would have made the geometry of the setup substantially more complex. For instance, various relative angles between the electron beam and laser pulse are considered in Refs.~\cite{vranic_sci_rep_2018, blackburn_njp_2019}.

\section{Photon emission and pair production}\label{sec:pairs}

In order to describe the QED processes leading to the production of positrons, we will employ the corresponding QED rates within the locally-constant field approximation (LCFA). Let us introduce the Lorentz-invariant quantum parameters $\eta = |F_{\mu \nu} p^\nu|/c^4$ and $\chi = |F_{\mu \nu} k^\nu|/(2c^4)$ characterizing the electron and photon dynamics in the external field, respectively. The field itself can be described by two Lorentz invariants $\mathcal{F}  = (\boldsymbol{E}^2 - \boldsymbol{B}^2)/E_\text{c}^2$ and $\mathcal{G}  = \boldsymbol{E} \cdot \boldsymbol{B}/E_\text{c}^2$, where $E_\text{c} = c^3 \approx 1.3\times 10^{16}~\text{V}/\text{cm}^2$ is the Schwinger critical field strength. The LCFA approach is based on the fact that if $\mathcal{F}$,~$\mathcal{G} \ll 1$ and $\mathcal{F}$,~$\mathcal{G} \ll \eta^2$,~$\chi^2$, then the QED rates are determined only by $\eta$ and $\chi$ given that $\mathcal{F}$ and $\mathcal{G}$ are small no matter what specific field configuration is examined~\cite{nikishov_ritus_1964_1}. This means that one can describe the necessary QED processes choosing the most convenient scenario. Basically, one considers either a constant crossed field~\cite{nikishov_ritus_1964_1, ritus_1985}, or a static magnetic field~\cite{erber_1966}. In the present study, we choose the latter configuration following the emission and pair production model employed in Refs.~\cite{kirk_2009, duclous_2011, ridgers_jcp_2014, arber_ppcf_2015}. Since for the laser intensities considered in our study, the field amplitude $E_0$ is always much smaller than the Schwinger limit, $E_0 \ll E_\text{c}$, and the dimensionless parameter $a_0 = E_0/(c\omega_0)$ obeys $a_0 \gg 1$, the LCFA is well justified within our simulations. This issue will be discussed in more detail in Sec.~\ref{sec:final_results}.

To incorporate the QED processes in PIC methods, one basically performs Monte-Carlo simulations governed by the relevant QED rates (see, e.g., Refs.~\cite{ridgers_jcp_2014, arber_ppcf_2015, pukhov_1999, fonseca_2002, nerush_prl_2011, elkina_2011, bastrakov_comp_2012, gonoskov_pre_2015, derouillat_2018}). However, in the present study, we compute the number of positrons produced according to the following scheme: (i)~we evolve the momentum and coordinates of a given electron portion taking into account the Lorentz force and the recoil force due to quantum radiation reaction; (ii)~at each time step, we also calculate the number density of photons emitted with a given quantum parameter $\chi$ which propagates then along a straight line; (iii)~traveling in the external field, this ``photon portion'' can decay via the nonlinear Breit-Wheeler process, so we sum the corresponding contributions to the positron number.
The main drawbacks of this approach are the following: (a)~it does not take into account the further stages of the cascading process, i.e., we consider only photons emitted by the primary (seed) particles; (b) we neglect the internal plasma field induced by the charged particles. These two aspects will be discussed in Sec.~\ref{sec:effects}. Moreover, this kind of a PIC simulation does not provide any information on the angular distribution of the positrons created. Given the threshold behavior of the scenario under consideration, the total number of particles already represents a relevant quantity which can be used for the intensity diagnostics. Besides, the advantage of our approach is that the first stage of the truncated cascade process is described without Monte-Carlo methods, so we perform only one run of our simulations for given laser field parameters. Let us turn to the detailed description of our approach.

\begin{widetext}

The $i$th electron portion, which has the weight $\Delta N^{(\text{el})}_i$, follows the classical equations of motion~\eqref{eq:eom} for $t \geq t_i$, i.e., we evaluate its momentum components and coordinates at each time step $t = t_j$, where $j \geq i$. At a given time instant $t = t_j$, the electron can emit photon with 4-momentum $k^\mu_\gamma$. In terms of the quantum parameters $\eta$ and $\chi$, the emission rate reads~\cite{erber_1966, sokolov_ternov}:
\begin{equation}
\frac{dN^{(\gamma)}}{d\chi dt} = \frac{\sqrt{3}c}{2\pi} \frac{\eta}{\gamma} \frac{F(\eta, \chi)}{\chi},
\label{eq:qed_emission}
\end{equation}
where $F(\eta, \chi) = F_x (\eta, 2 \chi/\eta)$ is the quantum synchrotron function,
\begin{equation}
F_x (\eta, x) = x^2 y K_{2/3} (y) + (1-x)y \int \limits_y^\infty dz K_{5/3} (z).
\label{eq:qed_Fx}
\end{equation}
Here $y \equiv 2x/[3\eta (1-x)]$ and $K_n (z)$ are modified Bessel functions of the second kind. One also assumes that $F_x (\eta, x) = 0$ for $x \geq 1$, i.e., $\chi$ does not exceed $\eta /2$. Note that a focused external field described in Sec.~\ref{sec:dynamics} is more favorable for pair production than a simple plane-wave background since the latter does not allow the electron to reach large values of the quantum parameter $\eta$. By solving the equations of motion in the case of a plane wave, one can indeed explicitly demonstrate that $\eta = |\boldsymbol{E}|/c^3$ throughout the interaction process no matter what energy the particle gains.

To evaluate the recoil force, which should be plugged into Eq.~\eqref{eq:eom}, we assume that the photon momentum $\boldsymbol{k}$ is parallel to that of the electron: $k^\mu_\gamma = (\omega_\gamma/c) \, (1, \boldsymbol{n})$ and $\boldsymbol{p} = |\boldsymbol{p}| \boldsymbol{n}$. Using the momentum conservation, we obtain
\begin{equation}
\boldsymbol{F}_\text{rec} = - \boldsymbol{n} \int \limits_0^{\eta/2} d\chi \, \frac{dN^{(\gamma)}}{d\chi dt} \, \frac{\omega_\gamma (\chi)}{c}.
\label{eq:qed_recoil}
\end{equation}
The quantum parameters $\eta$ and $\chi$ can be evaluated via
\begin{eqnarray}
\eta &=& \frac{1}{c^4} \sqrt{(\boldsymbol{E}p_0 + \boldsymbol{p} \times \boldsymbol{H})^2 - (\boldsymbol{E} \cdot \boldsymbol{p})^2},
\label{eq:qed_eta_chi_most_gen_1}\\
\chi &=& \frac{\omega_\gamma}{2c^5} \sqrt{(\boldsymbol{E} + \boldsymbol{n} \times \boldsymbol{H})^2 - (\boldsymbol{E} \cdot \boldsymbol{n})^2},
\label{eq:qed_eta_chi_most_gen_2}
\end{eqnarray}
where $p_0 = \gamma c$. The recoil force then reads
\begin{equation}
\boldsymbol{F}_\text{rec} = - \boldsymbol{n} \frac{\sqrt{3}c}{2\pi} \frac{\eta}{\gamma} \bigg [ \frac{(\boldsymbol{E}p_0 + \boldsymbol{p} \times \boldsymbol{H})^2 - (\boldsymbol{E} \cdot \boldsymbol{p})^2}{(\boldsymbol{E} + \boldsymbol{n} \times \boldsymbol{H})^2 - (\boldsymbol{E} \cdot \boldsymbol{n})^2} \bigg ]^{1/2} \int \limits_0^{1} dx \, F_x (\eta, x).
\label{eq:qed_recoil_most_gen}
\end{equation}
This force enters Eq.~\eqref{eq:eom} together with the classical Lorentz force. At each time step $t_j$, we also evaluate the number density of photons depending on $\chi$:
\begin{equation}
\frac{dN_{ij}^{(\gamma)}}{d\chi} = \frac{dN^{(\gamma)} (t_j)}{d\chi dt} \, \Delta t \Delta N^{(\text{el})}_i,
\label{eq:qed_ph_dens}
\end{equation}
where $i$ is the number of the electron portion and $\Delta t$ is the temporal grid step. These ``photon portions'' travel according to $\boldsymbol{r} (t) = \boldsymbol{r}_j + \boldsymbol{n}_j c (t - t_j)$ independently of $\chi$, where $\boldsymbol{r}_j$ and $\boldsymbol{n}_j$ are the electron's position and direction at $t = t_j$. Finally, we use the same temporal grid to evaluate the total number of positrons produced by the $j$th photon portions,
\begin{equation}
\Delta N^{(\text{pos})}_{ij} = \int \limits_0^{\eta/2} \! d \chi \! \int \limits_{t_j}^{+\infty} \! dt' \, \frac{dN_{ij}^{(\gamma)}}{d\chi} \, \frac{dN^{(\pm)}(t')}{dt'}.
\label{eq:qed_delta_n_pos_gen}
\end{equation}
Here $\eta$ is the quantum parameter of the electron at $t=t_j$. The upper limit of the $t'$ integration is practically the time instant when the photons escape from the external field. The QED rate of pair production can be represented as~\cite{erber_1966}:
\begin{equation}
\frac{dN^{(\pm)}(t')}{dt'} = \frac{c^3}{\omega_\gamma} \, \chi' T_\pm (\chi').
\label{eq:qed_PP_rate}
\end{equation}
Here $\omega_\gamma$ is the photon frequency determined by $\chi$, i.e. the quantum parameter at $t=t_j$, while $\chi'$ corresponds to the photon state at the time instant $t'$, i.e. $\chi' \equiv R^{1/2} (t') \chi$, where
\begin{equation}
R(t') = \frac{[(\boldsymbol{E}p_0 + \boldsymbol{p} \times \boldsymbol{H})^2 - (\boldsymbol{E} \cdot \boldsymbol{p})^2]~\text{at}~t'}{[(\boldsymbol{E}p_0 + \boldsymbol{p} \times \boldsymbol{H})^2 - (\boldsymbol{E} \cdot \boldsymbol{p})^2]~\text{at}~t_j}.
\label{eq:qed_delta_n_pos_ratio}
\end{equation}
The function $T_\pm$ has the form
\begin{equation}
T_\pm (\chi) = \frac{0.16}{\chi} \, K_{1/3}^2[2/(3\chi)],
\label{eq:qed_T_pm}
\end{equation}
so we arrive at
\begin{equation}
\Delta N^{(\text{pos})}_{ij} = \Delta t \Delta N^{(\text{el})}_i \frac{0.09\sqrt{3}}{2\pi c} \frac{\eta^2}{\gamma} \sqrt{(\boldsymbol{E}p_0 + \boldsymbol{p} \times \boldsymbol{H})^2 - (\boldsymbol{E} \cdot \boldsymbol{p})^2} \, \int \limits_0^{1} dx \, F_x(\eta, x) \!\! \int \limits_{t_j}^{+\infty} \! dt' \, R(t') \varkappa^2 \bigg ( \frac{4}{3 R^{1/2} (t') \eta x} \bigg ),
\label{eq:qed_delta_n_pos_final}
\end{equation}
where $\varkappa(z) \equiv z K_{1/3} (z)$. The square root containing the external fields in Eq.~\eqref{eq:qed_delta_n_pos_final} is evaluated at $t = t_j$.
Note that in this derivation we have not taken into account the fact that the number of positrons produced cannot exceed the initial weight of the photon. Since the pair production mechanism reduces the photon weight, one has to modify Eq.~\eqref{eq:qed_delta_n_pos_gen} according to
\begin{equation}
\Delta N^{(\text{pos})}_{ij} = \int \limits_0^{\eta/2} \! d \chi \, \frac{dN_{ij}^{(\gamma)}}{d\chi} \Bigg [ 1 - \mathrm{exp} \Bigg \{ - \!\! \int \limits_{t_j}^{+\infty} \! dt' \, \frac{dN^{(\pm)}(t')}{dt'} \Bigg \} \Bigg ].
\label{eq:qed_delta_n_pos_exp}
\end{equation}
This brings us to
\begin{equation}
\Delta N^{(\text{pos})}_{ij} = \Delta t \Delta N^{(\text{el})}_i \frac{\sqrt{3} c}{2\pi} \frac{\eta}{\gamma}  \, \int \limits_0^{1} dx \, \frac{F_x(\eta, x)}{x} \, \Big [ 1 - \mathrm{exp} \Big \{ - \!\! \mathcal{P} (\eta, t_j, x) \Big \} \Big ],
\label{eq:qed_delta_n_pos_exp_final}
\end{equation}
where
\begin{equation}
\mathcal{P} (\eta, t_j, x) = \frac{0.09 \eta x}{c^2} \, \sqrt{(\boldsymbol{E}p_0 + \boldsymbol{p} \times \boldsymbol{H})^2 - (\boldsymbol{E} \cdot \boldsymbol{p})^2} \! \int \limits_{t_j}^{+\infty} \! dt' R(t') \varkappa^2 \bigg ( \frac{4}{3 R^{1/2} (t') \eta x} \bigg ).
\label{eq:qed_delta_n_pos_exp_final_exp}
\end{equation}
The expression~\eqref{eq:qed_delta_n_pos_exp_final} is the leading-order contribution of Eq.~\eqref{eq:qed_delta_n_pos_exp_final} if the function~\eqref{eq:qed_delta_n_pos_exp_final_exp} is sufficiently small. The final expression~\eqref{eq:qed_delta_n_pos_exp_final} should be summed over the electron ($i$) and photon portions ($j$):
\begin{equation}
\Delta N^{(\text{pos})} = \sum_i \sum_{j > i} \Delta N^{(\text{pos})}_{ij}.
\label{eq:qed_delta_n_pos_sum_ij}
\end{equation}
The sum over $j$ involves all $t_j > t_i$. The quantity~\eqref{eq:qed_delta_n_pos_sum_ij} represents the number of positrons for a given initial position $\boldsymbol{r}_0$ of a Xe atom or free electron. It should then be summed over the spatial coordinates taking into account the number density of the atoms or free electrons depending on the scenario under consideration:
\begin{equation}
N^{(\text{pos})} = n \! \int \! d\boldsymbol{r}_0 \, \Delta N^{(\text{pos})},
\label{eq:final_volume}
\end{equation}
where either $n = n_\text{el}$ or $n = n_\text{Xe}$. We always assume that the number density of the Xe atoms is $n_\text{Xe} = 10^{14}~\text{cm}^{-3}$. Within the intensity domain under consideration, the external field usually ionizes 52 electrons of each atom, so we choose $n_\text{el} = 52 n_\text{el}$ to keep the number of the ``active'' electrons constant and focus on the enhancement due to more efficient dynamics of the particles in the case of xenon.

\end{widetext}

\section{Other effects and corrections}\label{sec:effects}

\subsection{Nuclear field versus laser field and the initial position of the electron ionized} \label{sec:nuclear_field}

Let us estimate the ratio $E_\text{nucl}/E$ at the time instant when the electron gets ionized and becomes a free particle according to our model. If the ionization energy is sufficiently large, the particle tunnels through the barrier and appears at some position $x_0$. To find this position, we can simply equate the potential energy of the particle in the combined field of the nucleus and laser with the ionization energy $I_p$. It brings us to
\begin{equation}
x_0 = \frac{I_p + \sqrt{I_p^2 - 4EZ}}{2E}
\label{eq:field_x0}
\end{equation}
We see that for $E > E_\text{BS} \equiv I_p^2/(4Z)$ the electron state is no longer classically bound [this value of $E$ exactly corresponds to the expression~\eqref{eq:F_BS}]. It is also convenient to introduce the maximal value of the particle's potential energy $-V_0 = -2\sqrt{ZE}$, so the condition $E = E_\text{BS}$ is equivalent to $I_p = V_0$.

The barrier-suppression regime relates to $\beta \equiv I_p/V_0 \leq 1$ and obviously appears when one considers the outer-shell electrons. However, as the major part of $e^+e^-$ pairs is produced when the inner-shell electrons emit high-energy photons, one has to evaluate first $\beta$ for these states. It turns out that even for the 52th electron of the Xe atom ($Z \approx 51$, $I_p = 9810.37$~eV), $\beta \approx 1.09$ if $I = 10^{22}~\text{W}/\text{cm}^2$ and $\beta \approx 0.61$ if $I = 10^{23}~\text{W}/\text{cm}^2$. This means that for such intense laser fields, the concept of $x_0$ [Eq.~\eqref{eq:field_x0}] is not well defined. To demonstrate that the ion field can be neglected once the particle is free, it suffices to compare the laser field to that of the nucleus for the initial (bound) electron state, i.e., instead of Eq.~\eqref{eq:field_x0}, we use the corresponding Bohr-orbit radius $r = \nu^2/Z$, so $E_\text{nucl} = Z/r^2 = Z^3/\nu^4 = \kappa^4/Z = 16 E_\text{BS}$. Accordingly, $E_\text{nucl}/E = 16 E_\text{BS}/E = 16 \beta^2$~\cite{artemenko_pra_2017}. For the $2s^2$ shell of Xe, this ratio obeys $E_\text{nucl}/E \lesssim 1$ if $I \gtrsim 10^{24}~\text{W}/\text{cm}^2$. Introducing a second pulse, one reduces the ratio $\beta^2$ by a factor of $2$, so we can safely neglect the nuclear field within the whole range $10^{23}~\text{W}/\text{cm}^2 \lesssim I \lesssim 10^{26}~\text{W}/\text{cm}^2$ for our order-of-magnitude estimates.

Next, we will estimate the role of the Coulomb field of the nuclei within the acceleration stage. In Eq.~\eqref{eq:eom}, we totally neglect the ion field $Z/r^2$. Let us first evaluate the distance where the field of the nucleus becomes small compared to the laser field:
\begin{equation}
\frac{Z}{r_0^2} = \frac{E}{100}.
\label{eq:field_r0}
\end{equation}
For $I = 10^{22}~\text{W}/\text{cm}^2$ and $Z = 54$ it leads to $r_0 \approx 3.2$~a.u. (for larger $I$ and smaller $Z$, $r_0$ is even smaller). We can estimate the total volume where the ion field is non-negligible according to $V_\text{ion} = (4/3) \pi r_0^3 N_\text{ion}$. For the relative volume factor, it brings us to
\begin{equation}
\frac{V_\text{ion}}{V} = \frac{4}{3} \pi r_0^3 n \approx 2\times10^{-9} \, (n \times 10^{-14}~\text{cm}^{3}),
\label{eq:field_volumes}
\end{equation}
which amounts to $2\times 10^{-7}$ even for $n = 10^{16}~\text{cm}^{-3}$. Here we do not take into account that the attractive force which the nucleus exerts on the electron can make the latter get closer to the field center, where the field is larger than $Z/r_0^2$. Nevertheless, since (a) the laser field direction is independent of the ion position, (b) we assumed that at $r=r_0$ the ion field is 100 times weaker than the laser field, (c) the electron is unlikely to be captured by the nucleus due to the small capture cross section, the ion field does not have any significant impact on the particle's trajectory. These rough estimates were also confirmed by numerical simulations.

Finally, we note that the distance $r$ (or $x_0$ for $\beta \geq 1$) is always much smaller than the characteristic length scale of the external field, which is $\lambda \sim 1~\mu\text{m}$. For example, for the $2s^2$ electrons $r \sim 10^{-6}~\mu\text{m}$, and for the outer-shell electrons $r \sim 10^{-5}~\mu\text{m}$. For weaker fields, e.g. $I = 10^{22}~\text{W}/\text{cm}^2$, we receive $x_0 \sim 10^{-5}~\mu\text{m}$. Therefore, we always assume that after ionization free electrons appear at the position of the nucleus.

\subsection{Ion motion within the ionization process} \label{sec:ion_motion}

Now we roughly estimate the effects of the ion motion. The maximal momentum appearing due to the ponderomotive forces is $p_\text{max} \sim EZ/\omega$. The relativistic parameter $\gamma$ has then the form $\gamma \approx (1 + [EZ/(Mc\omega)]^2)^{1/2}$. For Xe $Z \lesssim 50$, $M \approx 2.4 \times 10^5$~a.u., so even for $I = 10^{26}~\text{W}/\text{cm}^2$ we have $\gamma \lesssim 2.0$, which leads to $v_\text{max}/c \lesssim 0.87$ (for $I = 10^{24}~\text{W}/\text{cm}^2$ it yields $v_\text{max}/c \lesssim 0.17$). The maximal displacement of the ion can be estimated as $(\delta x)_\text{max}/\lambda \lesssim v_\text{max}/(4c)$, which never exceeds $0.22$ even for $I = 10^{26}~\text{W}/\text{cm}^2$. However, it is only a rough upper-bound estimate, so the ion motion is likely to be totally negligible even for such high intensities, provided one sums the results over the whole interaction region.

\subsection{Ion induced QED processes} \label{sec:ion_induced}

Although we are mainly interested in the two-stage process of pair production by a high-energy electron as discussed above, there are also several ion-induced phenomena that take place during the interaction. Namely, so far we have not addressed (a)~Bethe-Heitler pair production by a photon in the presence of the Coulomb field of the high-$Z$ nucleus, (b) bremsstrahlung. These ``collisional'' processes are in fact negligible according to the following estimates.

{\it Bethe-Heitler process}. Assuming that $\omega_\gamma \gg c^2$, $\alpha Z \ll 1$, and the energy of the electron/positron is much larger than $c^2$, one can show that the total cross section of the process in our units has the form~\cite{blp}: $\sigma_\text{BH} \approx (28/9) (1/c^5) Z^2 [\ln (2\omega_\gamma/c^2) - 109/42]$. The pair production rate then reads $dN_\text{BH}/dt = \sigma_\text{BH} c n$. This quantity should be multiplied by the number of high-energy photons in the interaction volume and by the characteristic photon lifetime $t_\gamma$: $N_\text{BH} = N_\gamma \sigma_\text{BH} c n t_\gamma$. For $\omega_\gamma = 1$~GeV, it yields $N_\text{BH} \approx 7.6 \times 10^{-19} \, N_\gamma Z^2 t_\gamma (n \times 10^{-14}~\text{cm}^{3})$. The number of high energy photons is very unlikely to exceed $10^3 \, N^{(\text{pos})}$, so for $Z=54$ one finds $N_\text{BH}/N^{(\text{pos})} < 2.2 \times 10^{-12} \, t_\gamma (n \times 10^{-14}~\text{cm}^{3})$. Assuming then $t_\gamma < 100 \lambda/c$, we arrive at $N_\text{BH}/N^{(\text{pos})} < 10^{-5}$ even for $n = 10^{16}~\text{cm}^{-3}$, so the Bethe-Heitler process provides a negligible contribution.

{\it Bremsstrahlung}. If both the initial and final energy of the electron are much larger than $c^2$, the cross section has the form $\sigma_\text{Br} \approx (2^{5/2}/3) (1/c^5) Z^2 \gamma_e^{1/2}$~\cite{artemenko_pra_2017}, where $\gamma_e$ is the corresponding Lorentz factor of the electron. The total number of photons can be estimated as $N_\text{Br} \approx N_e \sigma_\text{Br} c n t_e$, where $N_e$ is the characteristic number of high-energy electrons corresponding to $\gamma_e$ and $t_e$ is the characteristic interaction time of a high-energy electron. Accordingly, for $1$~GeV electrons ($\gamma_e \approx 2000$), $\sigma_\text{Br} \approx 1.7 \times 10^{-9} \, Z^2$ and $N_\text{Br} \approx 3.5 \times 10^{-18} \, N_e Z^2 t_e (n \times 10^{-14}~\text{cm}^{3})$. For the upper-bound estimation, we employ $Z = 54$, $t_e = 100 \lambda/c$, and $N_e = (10\lambda)^3 n \sim 10^7$. Therefore, $N_\text{Br} < 1.4\times 10^{-5} \, (n \times 10^{-14}~\text{cm}^{3})^2$, which amounts to $0.14$ even for $n = 10^{16}~\text{cm}^{-3}$. This means that bremsstrahlung can also be neglected.

Besides, we do not take into account further stages of the cascade process, when the electron/positron created emits one more photon leading to subsequent pair production. Such a scenario may give rise to an avalanche-like reaction, so one should either estimate the additional contributions or demonstrate that they are small compared to what we incorporate in Sec.~\ref{sec:pairs}. This issue will be discussed in the following subsection.

\begin{figure*}
\center{\includegraphics[height=0.35\linewidth]{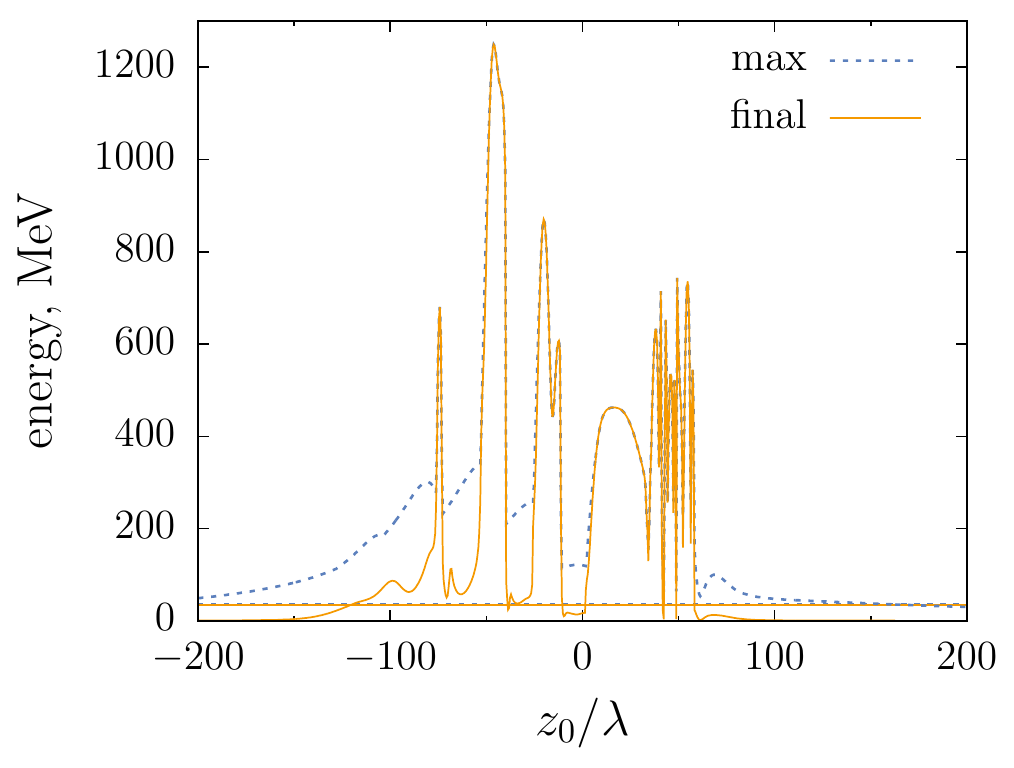}~~~~\includegraphics[height=0.35\linewidth]{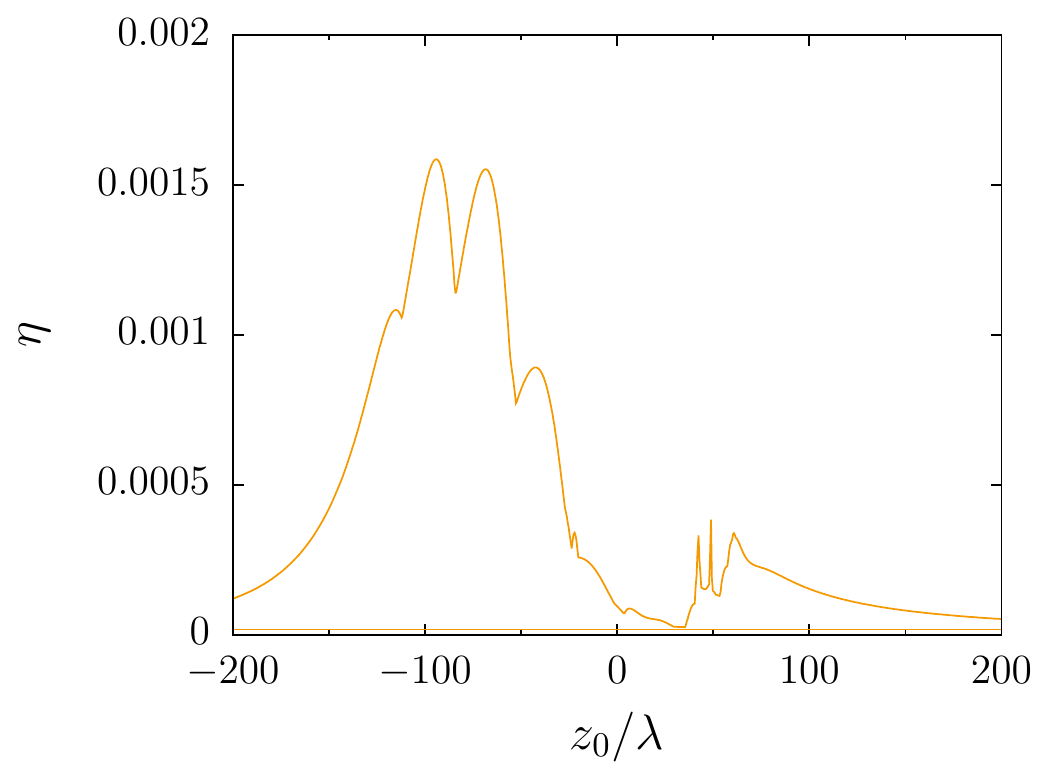}}
\caption{(left) Final and maximal energy of the electron and (right) maximal value of $\eta$ as a function of $z_0$ for $I = 2\times 10^{22}~\text{W}/\text{cm}^2$ in the case of an individual focused laser pulse ($x_0 = y_0 = 0$). The horizontal lines represent the corresponding values for the plane-wave case.}
\label{fig:en_eta_2}
\end{figure*}

\subsection{Further cascading} \label{sec:cascading}

Unlike the so-called S-type (``shower'') cascades where the $e^+e^-\gamma$ production processes basically consume energy from the seed high-energy particles, the phenomenon under consideration may correspond to A-type (``avalanche'') cascades (see, e.g., Refs.~\cite{fedotov_prl_2010, 2015_ufn_narozh}). The external laser field not only serves as a background for the Compton and Breit-Wheeler mechanisms but also accelerates particles, so that they regain the energy lost. If the external field is sufficiently strong, so the acceleration time $t_\text{acc}$ and the electron/photon lifetime $t_{e,\gamma}$ are much smaller than the interaction time $t_\text{esc}$, then a continuous avalanche-like reaction occurs being limited only by the laser pulse duration and ponderomotive expulsion of the particles.

As was demonstrated in Refs.~\cite{fedotov_prl_2010, 2015_ufn_narozh}, in the case of two laser pulses, the hierarchy $t_\text{acc} \ll t_{e,\gamma} \ll t_\text{esc}$ takes place when $\mu \equiv E_0/(\alpha E_\text{c}) \gtrsim 1$ and $\mu^{1/4} \gg (1/\alpha) \sqrt{\omega/c^2}$. For this scenario, we analyze the pulses with intensity up to $10^{24}~\text{W}/\text{cm}^2$, which yields $\mu = 0.28$ and also satisfies the second condition. It means that the enhancement of the positron number due to the further stages of cascading does not take place in our case. In the case of an individual focused laser pulse, an avalanche-like cascade can occur only when $\mu \gtrsim 1/\Delta$ for sufficiently large divergence of the beam, $\Delta > 0.03$~\cite{mironov_talk}. Since in our case $\Delta \approx 0.06$, we extent the intensity interval up to $10^{26}~\text{W}/\text{cm}^2$. We refrain from using higher intensities in our simulations as they could make the further stages of cascading non-negligible.

\section{Numerical results}\label{sec:results}

In this section, we will discuss the process of acceleration of electrons in the laser field and pair creation seeded by free electron gas and neutral Xe atoms. The external field will be chosen in the form of a single focused laser pulse described by Eqs.~\eqref{eq:field_Hx}--\eqref{eq:field_phi_tilde} and a combination of two counterpropagating laser pulses ($\theta = \pi$). The external field parameters are $\lambda = 1.0~\mu\text{m}$, $\rho_\text{F} = 2.0~\mu\text{m}$, $z_\text{F} = 20.3~\mu\text{m}$, $\mathfrak{F} = 51.4~\mu\text{m}$, $f_\# \equiv \mathfrak{F}/(2 \rho_0) = 10.0$, $\varphi = 0$, and $\tau = 33.3~\text{fs}$. The number of the carrier cycles is $N_\text{c} = 10$.

\subsection{Acceleration of particles}\label{sec:acc}

\begin{figure}
\center{\includegraphics[width=0.95\linewidth]{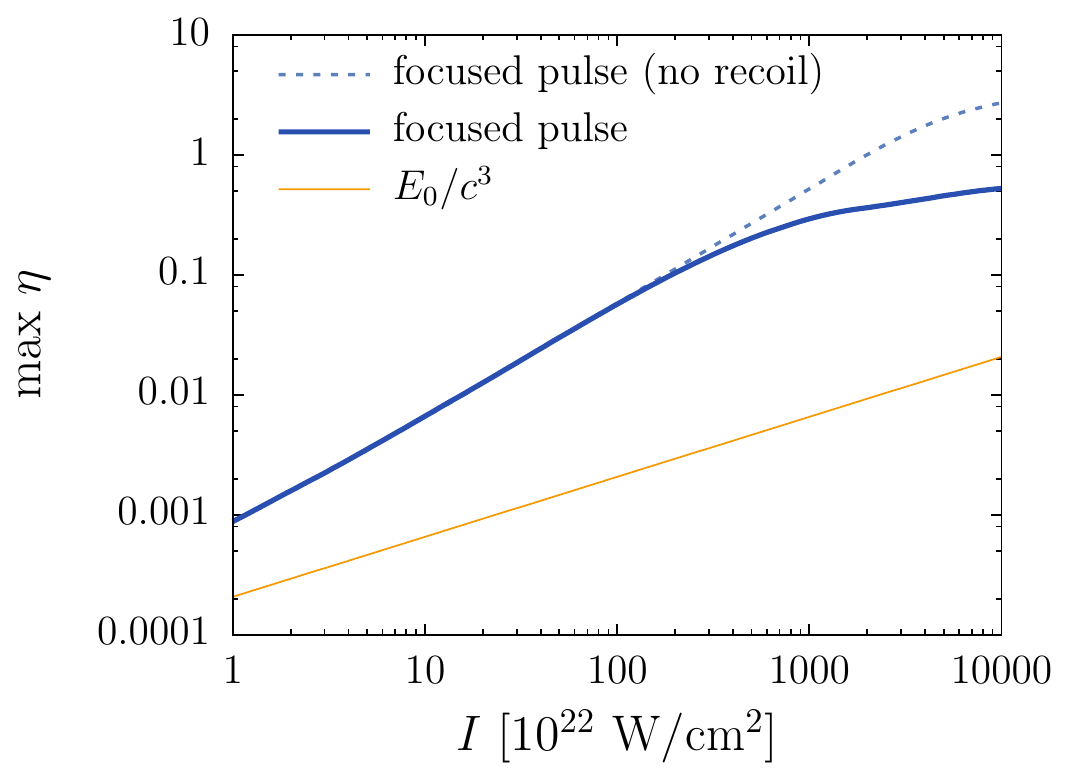}}
\caption{Maximal value of $\eta$ over time and $z_0$ as a function of $I$ in the case of a single focused pulse with the recoil force~\eqref{eq:qed_recoil_most_gen} (blue solid line) and without it (dashed line). The orange solid line displays the upper-bound estimate $\eta_\text{max} = E_0/c^3$ for the case of a plane-wave pulse ($x_0 = y_0 = 0$).}
\label{fig:eta_I0}
\end{figure}

First, we evolve the free electron's trajectory taking into account the recoil force~\eqref{eq:qed_recoil_most_gen} in order to find out what energies can be achieved when using (a) focused laser pulse, (b) simple plane-wave field. In Fig.~\ref{fig:en_eta_2} we depict the $z_0$ dependence of the final and maximal energy of the electron and the parameter $\eta$ for $I = 2\times 10^{22}~\text{W}/\text{cm}^2$ and $x_0 = y_0 = 0$. The electron is assumed to be free, i.e., no Coulomb forces are present. The horizontal lines correspond to the case of a finite plane-wave pulse (the results are $z_0$-independent). Note that in the case of a plane wave, the parameter $\eta$ is always equal to $|\boldsymbol{E}|/c^3$, i.e., it is determined by the local value of the electric field strength, which does not exceed the amplitude $E_0$. For instance, in Fig.~\ref{fig:en_eta_2}~(right) the maximal value of $\eta$ is $1.7 \times 10^{-5}$, while $E_0/c^3 = 2.9 \times 10^{-4}$. The difference appears due to the fact that the electron's trajectory does not cross any of the field maxima. On the other hand, the value of $\eta$ is crucial since positron production can only be efficient if the electrons emit high-energy photons. To further illustrate that the focused background is much more efficient in terms of reaching large $\eta$, we display the maximal (over $z_0$) values of $\eta$ as a function of $I$ together with the upper-bound estimate $\eta_\text{max} = E_0/c^3$ for the case of a plane-wave pulse (see Fig.~\ref{fig:eta_I0}).
\begin{figure*}[t]
\center{\includegraphics[height=0.35\linewidth]{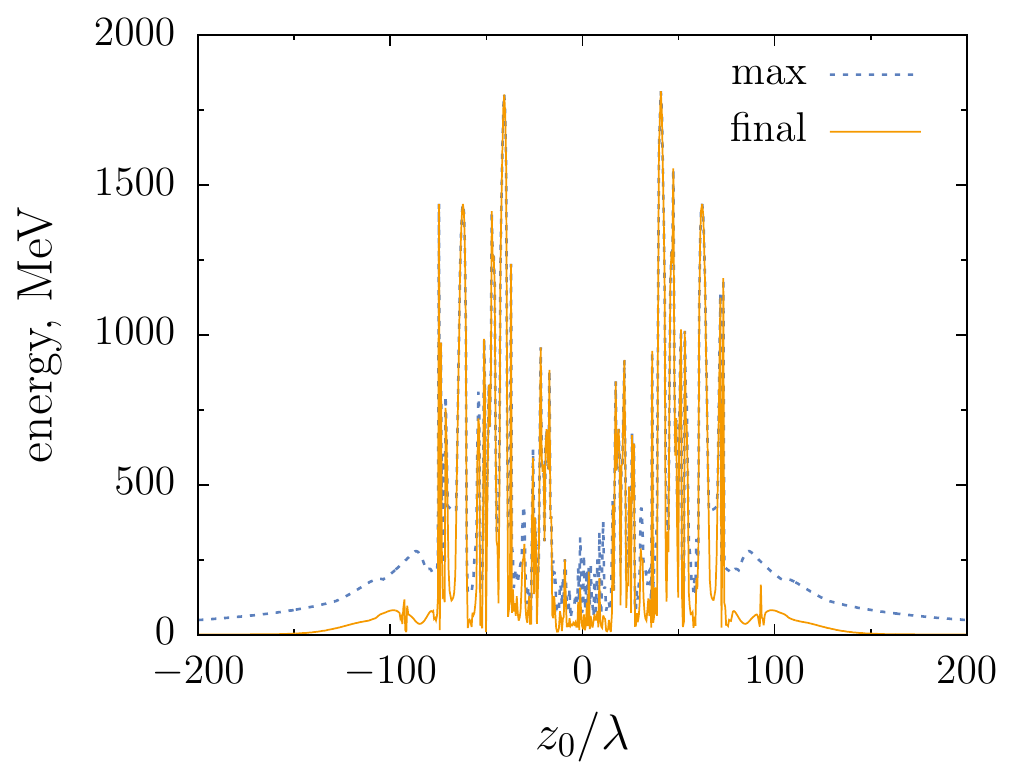}~~~~\includegraphics[height=0.35\linewidth]{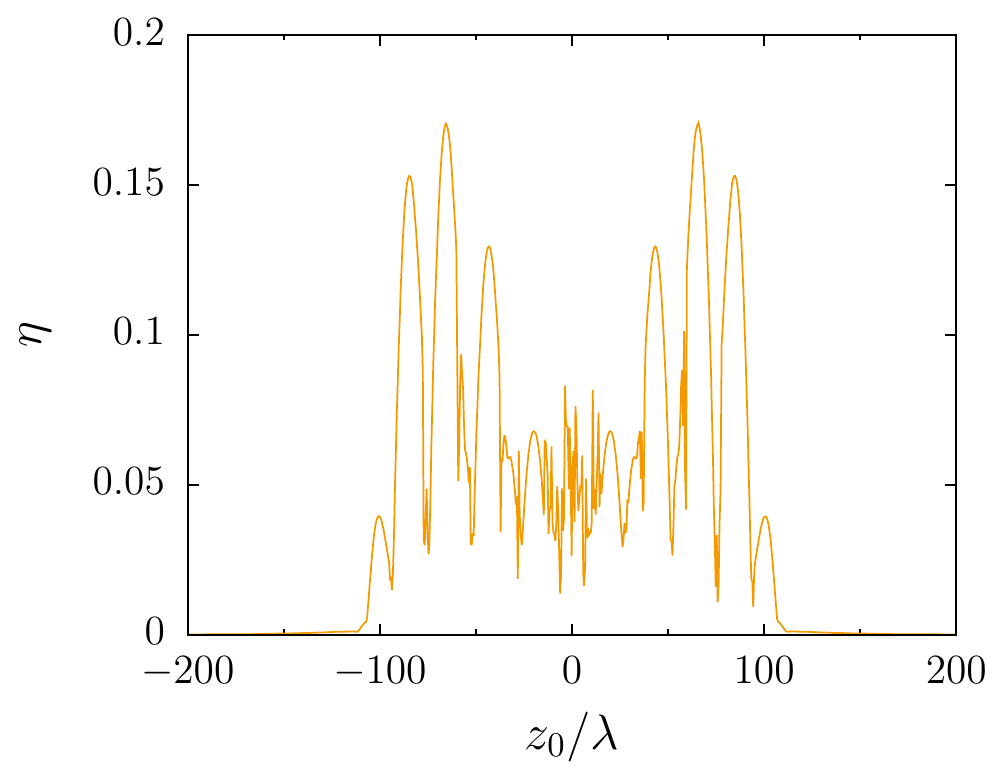}}
\caption{$z_0$ dependences of the maximal energy and $\eta$ in the case of free electrons traveling in the combination of two counterpropagating laser pulses with $I = 2 \times 10^{22}~\text{W}/\text{cm}^2$ ($x_0 = y_0 = 0$).}
\label{fig:en_eta_free_clp}
\end{figure*}
We observe that the results for a focused pulse are always much larger. Moreover, it may well be that the actual values of $\eta$ in the case of a plane-wave pulse are even smaller than $E_0/c^3$. For example, to make the plane-wave background more realistic, one can introduce a certain spatial cut-off function allowing the particles to escape from the interaction region. However, the results strongly depend on this profile, so we do not display them in Fig.~\ref{fig:eta_I0}. Since the field of a focused pulse is naturally finite in space, we will study this more realistic setup refraining from further discussions of the plane-wave background. In Fig.~\ref{fig:eta_I0} we also present the results obtained without taking into account the radiation reaction force~\eqref{eq:qed_recoil_most_gen} (dashed line). This force plays a notable role only when the laser intensity reaches $10^{24}$--$10^{25}$~$\text{W}/\text{cm}^2$. As will be demonstrated in what follows, the setup under consideration provides a non-negligible amount of pairs only when $I \gtrsim 10^{25}$~$\text{W}/\text{cm}^2$. Although the focused laser pulse indeed allows one to obtain large $\eta$, i.e. this scenario seems to be quite promising, the pair creation process is expected to have a quite high threshold.

To make the setup more efficient, we will introduce a second pulse propagating in the opposite direction. In Fig.~\ref{fig:en_eta_free_clp} we present the analogous $z_0$ dependences for the case of two focused counterpropagating laser pulses. Each of the pulses has the intensity $I = 2 \times 10^{22}~\text{W}/\text{cm}^2$. The presence of two pulses make the field configuration much more inhomogeneous and considerably more powerful in terms of the field strength. As a result, both the energy of the particles and parameter $\eta$ reach much larger values. Since this scenario is very far from the simplest plane-wave background not solely due to the focusing, in Fig.~\ref{fig:en_eta_free_clp}~(right) we observe the values which are about two orders of magnitude larger than those given in Fig.~\ref{fig:en_eta_2}~(right).

\begin{figure}
\center{\includegraphics[width=0.95\linewidth]{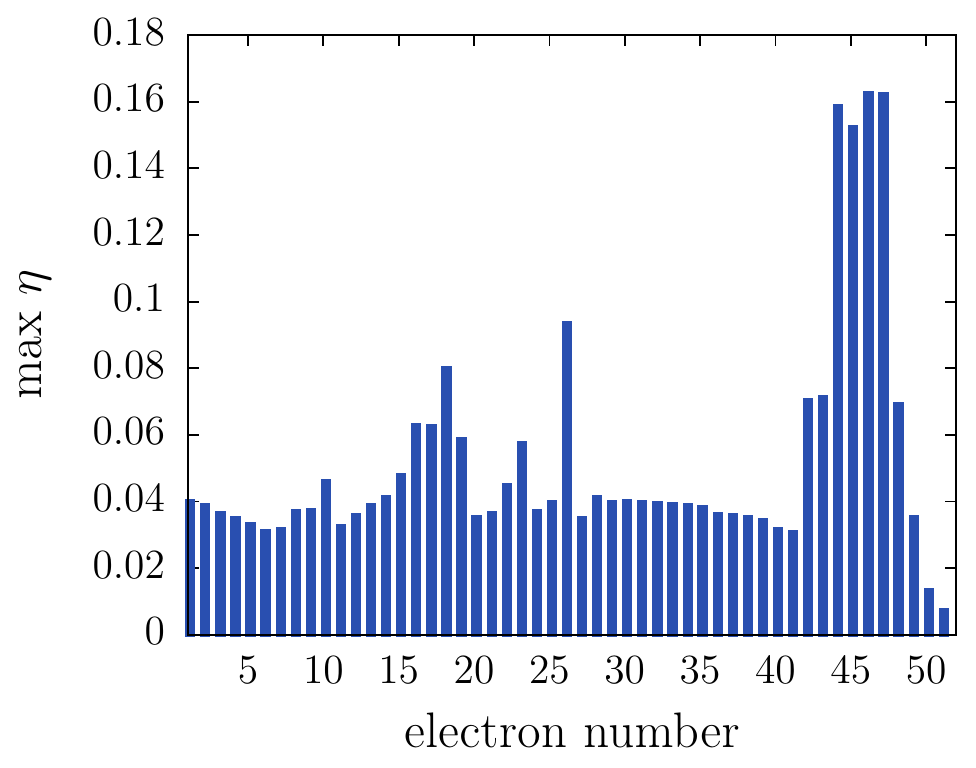}}
\caption{Maximal value of $\eta$ for various electrons of a Xe atom in the combination of two counterpropagating laser pulses ($I = 2 \times 10^{22}~\text{W}/\text{cm}^2$, $x_0 = y_0 = 0$, $z_0 = 5~\mu\text{m}$).}
\label{fig:histo}
\end{figure}

Finally, we take one more step in pursuit of reaching higher kinetic energies of the electrons by considering Xe atoms instead of free electrons. The advantage of this scheme consists in very efficient acceleration of the inner-shell electrons, which are being ionized only in a space-time region where the external laser field is sufficiently strong~\cite{maltsev_prl_2003, hu_pre_2006, karmakar_2007}. The Coulomb field of the nucleus preserve these electrons from premature ionization, so they gain relatively higher energies before they escape from the interaction region of the focused pulse(s). To illustrate this point, we present the histogram showing the maximal values of $\eta$ reached by different electrons of a Xe atom in the field of two counterpropagating laser pulses of the intensity $I = 2 \times 10^{22}~\text{W}/\text{cm}^2$ (see Fig.~\ref{fig:histo}). The two deepest inner-shell electrons (53rd and 54th) are not ionized at all, while the other 52 electrons are fully ionized. This plot reveals a typical situation where several electrons corresponding to a large ionization potential reach substantially larger values of $\eta$. Since this parameter plays then a crucial role in the two-stage process of pair production, Xe atoms represent a very efficient setup as will also be demonstrated in what follows.

\begin{figure*}[t]
\center{\includegraphics[height=0.35\linewidth]{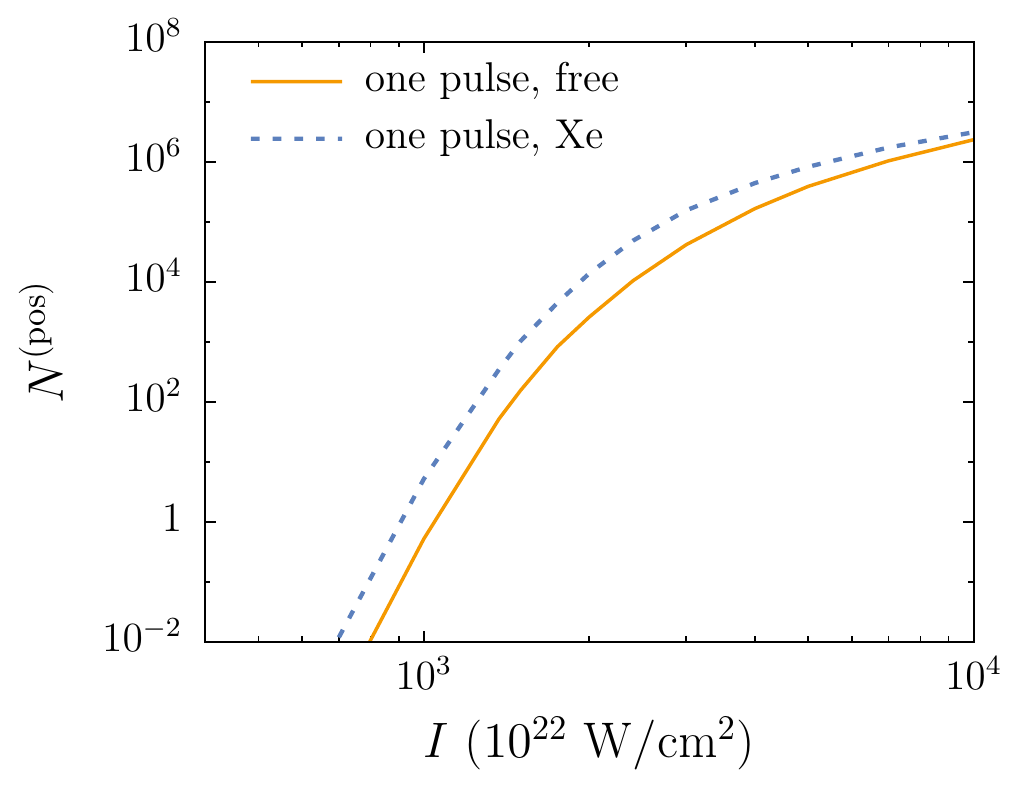}~~~~\includegraphics[height=0.35\linewidth]{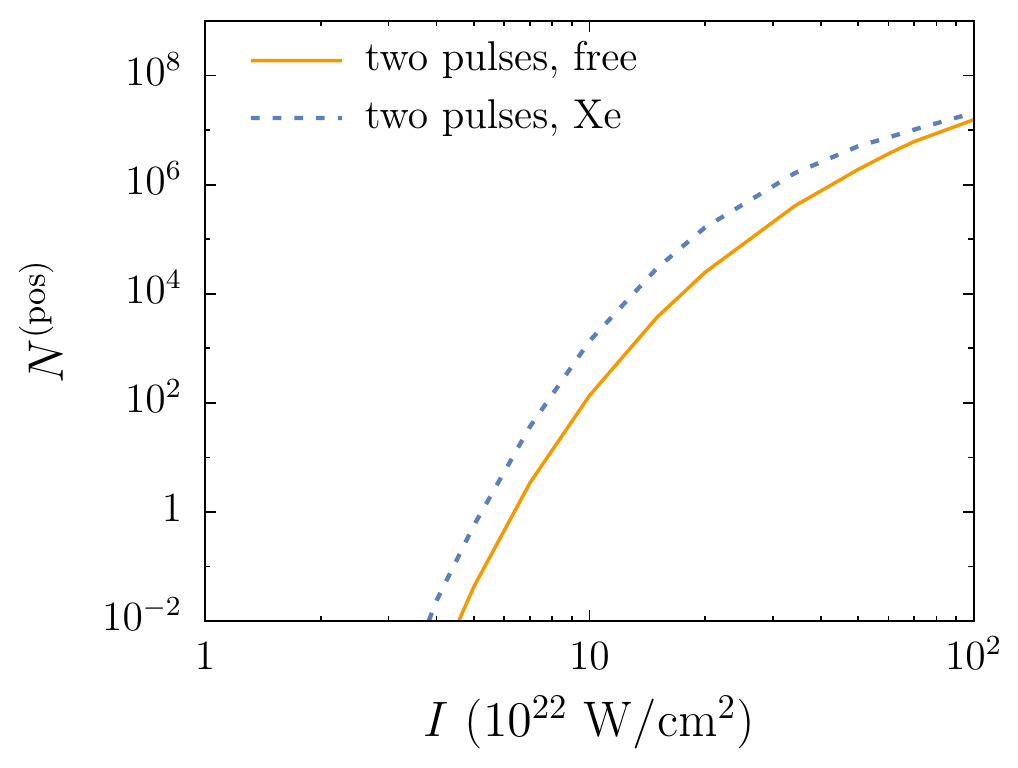}}
\caption{Total number of positrons produced in (left)~an individual laser pulse and (right) the combinations of two counterpropagating laser pulses as a function of the laser intensity $I$. The seed particles are either free electrons (solid lines) or neutral Xe atoms (dashed lines).}
\label{fig:final_curves}
\end{figure*}

\subsection{Positron production}\label{sec:positrons}

In this section, we perform the calculations of the total number of positrons according to Sec.~\ref{sec:pairs} as a function of the laser intensity $I$. We analyze the process in the case of free Xe atoms randomly distributed over space with the number density $n_\text{Xe} = 10^{14}~\text{cm}^{-3}$ and free electrons with the density $n_\text{el} = 52 n_\text{Xe}$. As was shown in, e.g., Figs.~\ref{fig:en_eta_free_clp}, one has to employ a very high spatial resolution in terms of the initial position $\boldsymbol{r}_0$ to accurately describe the electron dynamics, i.e., accurately integrate in Eq.~\eqref{eq:final_volume}. Since the pair-production mechanism is highly nonlinear, analyzing our data , we decided to associate a quite large uncertainty with the spatial integration assuming that our numerical simulations provide only order-of-magnitude estimates. On the other hand, as will be shown further, this relatively low precision already allows one to recover the field intensity quite accurately. This will be discussed in more detail in the next subsection.

Before integrating over the initial spatial coordinates in Eq.~\eqref{eq:final_volume}, one has to sum over the electron and photon portions in Eq.~\eqref{eq:qed_delta_n_pos_sum_ij}. Unfortunately, this stage is very time consuming in the case of Xe atoms, where there are many electron portions, whereas a free electron is considered as only one macroparticle. To save the computational time, we estimate the positron yield in the case of Xe by using the maximal values of the $\eta$ parameter omitting the positron production stage described by Eqs.~\eqref{eq:qed_ph_dens}--\eqref{eq:qed_delta_n_pos_sum_ij}. Namely, we construct the dependence $\Delta N^{(\text{pos})} (\eta)$ analyzing the data obtained in the case of free electrons. Then we use this correspondence to estimate $\Delta N^{(\text{pos})}$ in the case of Xe atoms and integrate the results over $\boldsymbol{r}_0$ for each of the Xe electrons. Accordingly, we obtain 54 contributions which should then be summed up and multiplied with the number density $n_\text{Xe}$ (basically, there are 52 nonzero contributions). Taking into account that there is a significant ambiguity in $\Delta N^{(\text{pos})} (\eta)$, we estimated the additional uncertainty due to the simplified treatment of the Xe electrons and found out that the relative error in the positron yield is always considerably smaller than the order-of-magnitude error bars that we had introduced before.

The results are summarized in Fig.~\ref{fig:final_curves}. First, we observe that the pair-production threshold is indeed substantially lower in the case of two counterpropagating laser pulses, which can be used to detect positrons already for $I \gtrsim 10^{23}~\text{W}/\text{cm}^2$, while an individual focused pulse should be two orders of magnitude stronger ($I \gtrsim 10^{25}~\text{W}/\text{cm}^2$). Second, the two configurations considered in our study do not allow one to determine the laser intensity in the range $10^{24}~\text{W}/\text{cm}^2 \lesssim I \lesssim 10^{25}~\text{W}/\text{cm}^2$ for three reasons: (a) an individual pulse does not yet produce pairs, (b) we do not have reliable data in the case of two pulses as we neglect QED cascading, (c) even if the further stages of cascading were taken into account, the uncertainty in the laser intensity would be quite large because for $I \gtrsim 10^{24}~\text{W}/\text{cm}^2$ the positron number does not increase that dramatically with increasing $I$. Nevertheless, one can properly adjust the field configuration by changing the angle $\theta$ between two laser pulses allowing one to efficiently cover the whole interval $10^{23}~\text{W}/\text{cm}^2 \lesssim I \lesssim 10^{26}~\text{W}/\text{cm}^2$ and also minimize the uncertainties. This is basically the main statement of our study.

Third, in Fig.~\ref{fig:final_curves} it is demonstrated that using neutral Xe atoms instead of free electrons even with a lower number density notably enhances the effect of positron production. Namely, the results are several times larger than those obtained for free particles. For larger values of the laser intensity, the enhancement becomes not that evident since the external field immediately ionizes a major part of electrons making this setup quite similar to the initial state containing free electrons. Let us compare our results with those of Artemenko and Kostyukov~\cite{artemenko_pra_2017} for the case of a focused standing wave. First, we employed the data given in Ref.~\cite{artemenko_pra_2017} for He atoms dividing it by $2$ since the He atom contains two electrons which get ionized almost immediately due to the small ionization potential. The results were also normalized by taking into account the different values of the electron/ion density. Second, we compared the results for Xe atoms. In both cases our predictions were about one order of magnitude smaller indicating almost the same enhancement when using atoms instead of free electrons. Since the positron yield strongly depends on the field amplitude and the field configurations considered here and in Ref.~\cite{artemenko_pra_2017} are quantitatively different, we find a good agreement between our approach and that used in Ref~\cite{artemenko_pra_2017}.

In the present study we compute only the total number of positrons produced without specifying their angular distribution. In principle, this information can also be obtained by using more involved QED expressions. For instance, the energy distribution between the electron and positron produced via the Breit-Wheeler process was described in Ref.~\cite{daugherty_1983}). Such improvements require more time-consuming simulations and can be a subject of our future investigations.


\subsection{Intensity diagnostics and uncertainties}\label{sec:final_results}

Let us now discuss how the uncertainties of the positron yield can be estimated. To determine the accuracy of the ionization model employed, we perform the full calculations of the positron number replacing the ADK expressions with the PPT one. To take into account the possible inadequacy of the tunnel picture itself, we vary the parameter $\overline{\alpha}$ in Eq.~\eqref{eq:ioniz_prob_full} and the function $W_\text{L} (t)$ by $\pm 10\%$. The resulting changes were found to be small compared to the ADK--PPT discrepancy.

Next, we analyze the accuracy of the external-field model. First, Eqs.~\eqref{eq:field_Hx}--\eqref{eq:field_phi_tilde} are valid only when the paraxial approximation is well justified, i.e. $k_0 \rho_\text{F} \gg 1$. In our case $k_0 \rho_\text{F} \approx 12$. Second, although we propose a method to measure the laser peak intensity, the other parameters of the external field are not {\it a priori} known in the real experimental setups. For instance, the focal spot radius and the tightness of laser focusing are difficult to control. However, we do not assign any uncertainty to these parameters keeping in mind that it should be estimated later when the actual experimental setup is fully specified. Our numerical procedures can be applied within a wide domain of the field parameters, i.e., it is not limited just to the illustrative configuration analyzed in the present study.

Another important source of systematic errors is the Coulomb interaction among the particles. Here we assume that the main contribution appears due to the interaction between a given electron and the nuclei. To estimate the corresponding discrepancy, we perform the calculations assuming that the nuclei are randomly distributed over space with given average density and taking into account the Coulomb force. The ion motion in the laser field and the electron tunneling distance can be then completely neglected (see Sec.~\ref{sec:effects}).

We note then that the QED expressions~\eqref{eq:qed_emission} and \eqref{eq:qed_PP_rate} are valid in our case to a high accuracy because (a) the external field is much weaker than the Schwinger limit $E_\text{c} \approx 1.3 \times 10^{16}~\text{V}/\text{cm}$, (b) the parameters $\eta^2$ and $\chi^2$ are much larger than the Lorentz invariants of the external field, (c) the dimensionless field strength $a_0 \gtrsim 200$ is much larger than unity~\cite{ridgers_jcp_2014, arber_ppcf_2015}. We also note that in the case of the nonlinear Compton scattering, the locally constant field approximation underlying the expression~\eqref{eq:qed_emission} usually either is very accurate or fail in the region of low photon energy~\cite{di_piazza_pra_2019, ilderton_pra_2019}. Nevertheless, the subsequent Breit-Wheeler process is likely to occur only when the photon energy is high. A detailed analysis of the validity of the LCFA in the context of various QED processes can be found in recent papers~\cite{meuren_prd_2016, di_piazza_pra_2019, blackburn_2018, ilderton_pra_2019, aleksandrov_prd_2019_1, sevostyanov} and references therein. 

Besides the systematic errors, we should also take into account the possible numerical errors which may arise if the results do not fully converge. We make sure that the various nonphysical grid steps are sufficiently small, so that the corresponding inaccuracy is completely negligible. On the other hand, a large discrepancy is associated with the procedure of summing over the initial coordinate $\boldsymbol{r}_0$, which is quite complicated in three dimensions. By analyzing the data, we decided to assume that the integration provides only order-of-magnitude estimates. As a result, all of the other discrepancies discussed above become inessential.

\begin{figure}[t]
\center{\includegraphics[width=0.95\linewidth]{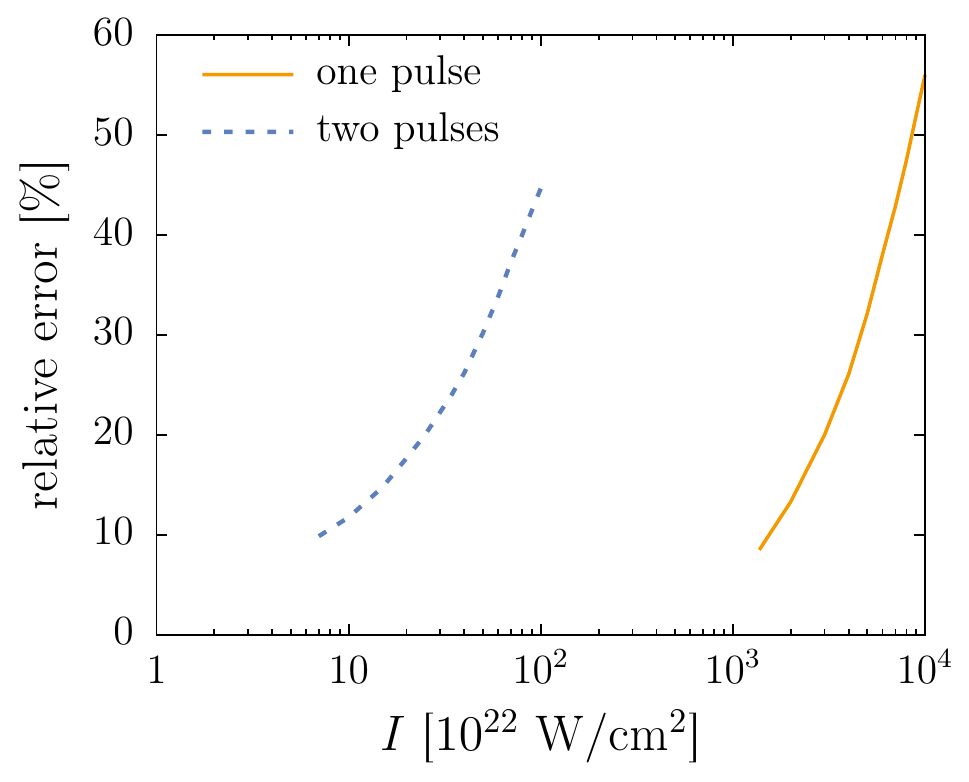}}
\caption{Relative uncertainty in the laser intensity $I$ as a function of $I$ in the case of one laser pulse (solid line) and the combination of two pulses (dashed line). For each of the two scenarios, the left edge of the intensity domain corresponds to a small positron yield $N^\text{(pos)}<1$, while the upper boundary appears since we disregard the cascading processes.}
\label{fig:final_dI}
\end{figure}

To determine the laser intensity, we make use of the sharp threshold behavior of the pair production process revealed in Fig.~\ref{fig:final_curves}. Indeed, if the number of positrons created is of the order $1$--$10^4$, even a notable uncertainty in $N^\text{(pos)}$ does not affect much the intensity. To present a transparent quantitative analysis of the corresponding uncertainty in the laser intensity, we calculate the relative error according to
\begin{equation}
\frac{\delta I}{I} = \frac{dI}{dN^\text{(pos)}} \frac{\delta N^\text{(pos)} (I)}{I},
\label{eq:I_error}
\end{equation}
where the derivative is computed using the numerical data and $\delta N (I)$ is chosen as indicated above. For a given scenario, increasing $I$, one receives larger values of $dI/dN$ as was mentioned above. We also underline that our simulations are performed only up to $10^{24}~\text{W}/\text{cm}^2$ in the case of two laser pulses and $10^{26}~\text{W}/\text{cm}^2$ in the case of a single pulse since we do not take into account the cascading process. On the other hand, for smaller $I$ the uncertainty $\delta N (I)$ becomes substantial and the number of positrons itself becomes tiny, so the domain where one can carry out the intensity diagnostics is, of course, limited from both sides. We summarize the uncertainties in Fig.~\ref{fig:final_dI} where the relative error in the laser intensity $I$ is depicted as a function of $I$. It turned out that the uncertainties do not depend on the choice of seed particles (free electrons of xenon). Choosing the appropriate scenario, one can perform a more accurate diagnostics. Using several setups in combination, one can obtain more reliable and precise values of the laser intensity.


\section{Conclusion} \label{sec:discussion}

In the present study, we provided quantitative estimates for the total positron yield in several experimental scenarios, which can be used as a tool for the laser intensity diagnostics. Namely, we examined the setups involving an individual laser pulse and a combination of two counterpropagating pulses. As seed particles we considered free electrons and Xe neutral atoms randomly distributed over space. The laser field accelerates the seed electrons which can subsequently emit high-energy photons. These photons can, in turn, decay via the Breit-Wheeler mechanism producing $e^+e^-$ pairs. Since this process has a sharp threshold dependence on the laser intensity, measuring the positron yield allows one to accurately extract the intensity when working in the vicinity of the pair production onset.

In order to calculate the number of positrons produced, we evolved the electron trajectories taking into account possible photon emission via the nonlinear Compton process and the Breit-Wheeler mechanism described by local expressions derived within QED. Neglecting the further cascading stages involving the secondary particles, we computed the total number of pairs summing them over the initial coordinates of the electrons or Xe atoms.

Different scenarios provide different working regimes, i.e. ranges of the laser intensity where it can be accurately determined. The corresponding domains cover a wide interval $10^{23}$--$10^{26}~\text{W}/\text{cm}^2$. According to our numerical results, choosing the appropriate setup ensures that the relative uncertainty in the laser intensity does not exceed $10$--$50\%$.

Although our estimates have already indicated that the laser intensity can be extracted with a high accuracy, there are several improvements that we aim to carry out within our future studies. First, one can examine some other experimental setups involving other field configurations, e.g., two laser pulses crossing at some angle $\theta \in (0, \pi)$, or other atoms as a source of seed electrons. Second, it is also desirable to predict the angular distribution of the positrons produced. Finally, we point out that in order to reduce the overall uncertainty, one can also take into account the further cascading stages, which become important at higher intensities, and refine the ionization model, so that it yields more accurate predictions for the ionization probabilities.


\end{document}